\documentclass[10pt,conference,a4paper]{IEEEtran}
\usepackage{steinmetz}
\usepackage{amssymb}
\usepackage[per-mode=symbol]{siunitx}
\usepackage{subcaption}
\usepackage{graphicx} 
\usepackage{url}
\usepackage{float}
\usepackage{cite}
\usepackage{pifont}
\usepackage{amsfonts}
\usepackage{multirow}
\usepackage{balance}
\usepackage{comment}
\usepackage{epstopdf}
\usepackage{booktabs}
\usepackage{tabularx}
\usepackage{xspace}
\usepackage{paralist}
\usepackage{mathrsfs}
\usepackage{multicol}
\usepackage{stmaryrd}
\usepackage{amsmath}
\usepackage{booktabs}
\usepackage{enumitem}
\usepackage{tikz}
\usetikzlibrary{calc}
\usepackage{pgfplots}
\pgfplotsset{compat=1.18} 
\usepackage{fancyhdr}
\def\bA{{\mathbf{A}}}    
\def\bF{{\mathbf{F}}}  \def\bH{{\mathbf{H}}} \def\bI{{\mathbf{I}}}

\def\bU{{\mathbf{U}}} \def\bV{{\mathbf{V}}} \def\bW{{\mathbf{W}}}  

\def\ba{{\mathbf{a}}} \def\bb{{\mathbf{b}}}   
\def\bf{{\mathbf{f}}}    
    
\def\bp{{\mathbf{p}}}    
  \def\bw{{\mathbf{w}}}  

\fancypagestyle{FirstPage}{
\lfoot{\copyright~2022 IEEE \hfill}
}

\begin{document}

\title{Unleashing 3D Connectivity in Beyond 5G Networks with Reconfigurable Intelligent Surfaces}


\author{\IEEEauthorblockN{Jiguang He$^1$, Aymen Fakhreddine$^{1,2}$, Arthur S. de Sena$^{3}$, Yu Tian$^1$, and Merouane Debbah$^{4}$}
\IEEEauthorblockA{$^1$Technology Innovation Institute, Abu Dhabi, United Arab Emirates}
\IEEEauthorblockA{$^2$Institute of Networked and Embedded Systems, University of Klagenfurt, Klagenfurt, Austria}
\IEEEauthorblockA{$^3$Centre for Wireless Communications, FI-90014, University of Oulu, Finland}
\IEEEauthorblockA{$^4$Khalifa University of Science and Technology, Abu Dhabi, UAE}
\IEEEauthorblockA{E-mails: \{jiguang.he@tii.ae, aymen.fakhreddine@tii.ae, arthur.sena@oulu.fi, yu.tian@tii.ae, merouane.debbah@ku.ac.ae\}
}}


\maketitle

\thispagestyle{empty}
\pagestyle{empty}
            




\begin{abstract}
Reconfigurable intelligent surfaces (RISs) bring various benefits to the current and upcoming wireless networks, including enhanced spectrum and energy efficiency, soft handover, transmission reliability, and even localization accuracy. These remarkable improvements result from the reconfigurability, programmability, and adaptation capabilities of RISs for fine-tuning radio propagation environments, which can be realized in a cost- and energy-efficient manner. In this paper, we focus on the upgrade of the existing fifth-generation (5G) cellular network with the introduction of an RIS owning a full-dimensional uniform planar array structure for unleashing advanced three-dimensional connectivity. The deployed RIS is exploited for serving unmanned aerial vehicles (UAVs) flying in the sky with ultra-high data rate, a challenging task to be achieved with conventional base stations (BSs) that are designed mainly to serve ground users. By taking into account the line-of-sight probability for the RIS-UAV and BS-UAV links, we formulate the average achievable rate, analyze the effect of environmental parameters, and make insightful performance comparisons. Simulation results show that the deployment of RISs can bring impressive gains and significantly outperform conventional RIS-free 5G networks.

\end{abstract}


\section{Introduction}
Drones, also known as unmanned aerial vehicles (UAVs), have recently undergone a tremendous expansion, generating a wide range of emerging applications, such as goods delivery, urban air taxis, remote surveillance, border control, agricultural or industrial monitoring, and disaster relief~\cite{Zhang2019}. Even though the aforementioned applications span distinct domains, the commonality among them is the demanding need for three-dimensional (3D) wireless connectivity for the transfer of real-time sensor data and control commands. Such connectivity must be reliable, secure, and supports high data rates, up to several hundred megabits per second (Mbps).

The current commercial fifth generation (5G) base stations (BSs) are customized with the primary purpose to provide a two-dimensional (2D) coverage to users on the ground. Providing ubiquitous three-dimensional (3D) coverage and even full earth coverage is a hot topic for the next generations of wireless networks, i.e., beyond 5G and sixth generation (6G)~\cite{Zhang2019_VTM}. To mitigate the lack of a dedicated infrastructure to serve flying UAVs, approaches that do not require substantial hardware upgrades or additional deployments to be borne by mobile network operators should be adopted. In other words, one should heavily rely on the existing 5G cellular networks and focus on its upgrade to meet the quality-of-service (QoS) requirements of UAV communications. To the best of our knowledge, the adoption of reconfigurable intelligent surfaces (RISs) to extend the 3D coverage of cellular networks for serving aerial users has not yet been fully explored.

In the recent literature, the works in \cite{lu2021aerial,mu2021intelligent,liu2020machine,li2020reconfigurable} focus on how the RIS can be applied to enhance UAV-enabled wireless networks. However, their purpose is also to serve ground users. To be specific, a terrestrial BS is able to communicate with distant ground users in the absence of line-of-sight (LoS) links by mounting an RIS on a UAV, which enables intelligent reflection from the sky~\cite{lu2021aerial}. In~\cite{mu2021intelligent,liu2020machine,li2020reconfigurable}, UAVs act as aerial BSs with flexible deployment. The authors in \cite{liu2021reconfigurable} dedicated a whole section reviewing UAV and RIS, and listed relevant publications that explored UAV-mounted RISs. 

In this paper, we leverage the reconfigurability capabilities of RISs to enhance the existing cellular networks by beamforming toward the sky to provide 3D connectivity to the UAVs. This offers an ``easy to implement" upgrade to the existing 5G cellular infrastructure, rendering beyond 5G (B5G), and enables an improved cellular architecture that incorporates large RISs in key locations at altitudes slightly lower than those of the serving BS antennas of currently deployed cellular networks. In this sense, unlike the existing works that consider RIS-mounted UAVs as part of the network either as relays or aerial BSs, our study considers the UAVs that are in fact cellular user equipment (UE) served by the network. Moreover, we take into account the LoS probability and study the average achievable rate under different propagation environments. Simulation results illustrate how the RIS-assisted B5G networks outperform their RIS-free counterparts in terms of average achievable rate and sky coverage.   

\textit{Notations}: A bold lowercase letter $\ba$ denotes a vector, and a bold capital letter $\bA$ denotes a matrix. $(\cdot)^\mathsf{T}$ and $(\cdot)^\mathsf{H}$ denote the matrix or vector transpose and Hermitian transpose, respectively. $\mathrm{Tr}(\cdot)$ denotes the trace operator, $\mathrm{diag}(\ba)$ and $ \mathrm{det}(\cdot)$ denote a diagonal matrix with the entries of $\ba$ on its diagonal and the determinant of a matrix, $\phase{a}$ returns the phase of the complex scalar $a$,
$\ba \otimes \bb$ denotes the Kronecker product of $\ba$ and $\bb$, $\bI_{M}$ denotes the $M\times M$ identity matrix, $\mathbf{0}$ is an all-zero matrix, $j = \sqrt{-1}$, and $\|\cdot\|_2$ denotes the Euclidean norm of a vector. $[\ba]_i$ and $[\bA]_{ij}$ denote the $i$th element of $\ba$ and the $(i,j)$th element of $\bA$, respectively. Finally, $|\cdot|$ returns the absolute value of a complex number.

\section{System model}\label{sec:sysmodel}
\begin{figure}[t]
	\centering
\includegraphics[width=0.9\linewidth]{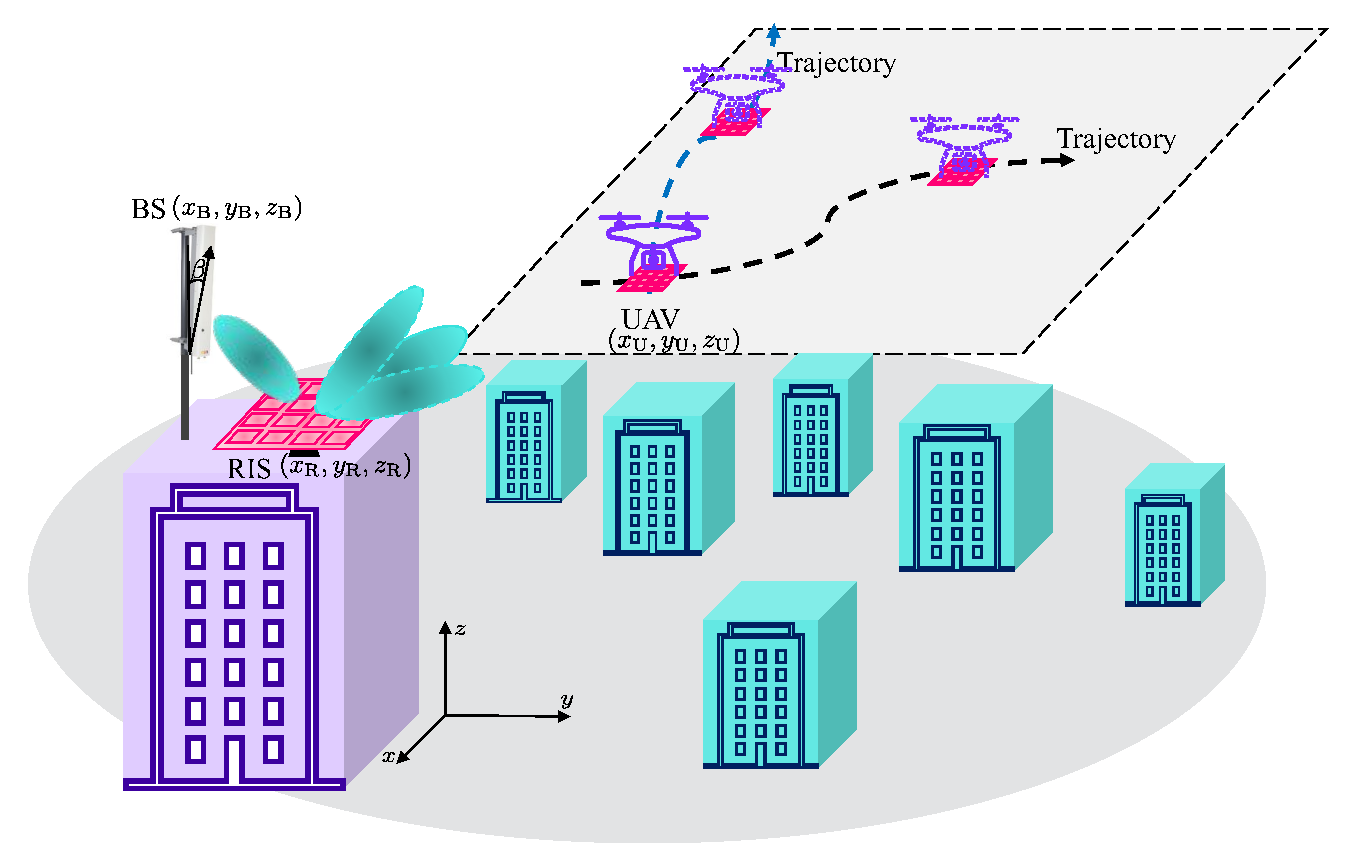}
	\caption{3D connectivity enabled by an RIS along with a legacy 5G BS. 
 }
		\label{System_Model}
		\vspace{-0.5cm}
\end{figure}

We focus on the 3D connectivity for UAV communications with the aid of a static RIS along with a legacy 5G BS, depicted in Fig.~\ref{System_Model}. The BS is equipped with a uniform linear array (ULA) structure, consisting of $N_\text{B}$ vertical antenna elements, down-tilted by a clockwise rotation of $\beta$ radians. To offer 3D connectivity for the flying UAV without additional deployment of costly next-generation BSs, we rely on the cost-efficient deployment of an RIS, which is capable of performing 3D beamforming thanks to its massive number of sub-wavelength meta-atoms. The RIS can generate a group of candidate beams to cover the whole sky, where the UAV is supposed to be located. Specifically, one RIS composed of $N_\text{R} = N_{\text{R},x} N_{\text{R},y}$ meta-atoms is deployed in the close proximity of the BS with LoS availability~\cite{Emil2022}, owning a uniform planar array (UPA) structure, which is parallel to the $x\text{-}y$ plane; $N_{\text{R},x}$ and $N_{\text{R},y}$ denote the number of RIS meta-atoms across $x$ and $y$ axes, respectively. The flying UAV is also supposed to employ a UPA with $N_\text{U} =N_{\text{U},x}N_{\text{U},y} $ antennas, where $N_{\text{U},x}$ and $N_{\text{U},y}$ are the number of antennas across $x$ and $y$ axes, respectively. We assume that the antenna planes of RIS and UAV are parallel to each other. The coordinates of the BS, the RIS, and the UAV are $\bp_\text{B}=(x_\text{B},y_\text{B},z_\text{B})^\mathsf{T} \in \mathbb{R}^{3}$,  $\bp_\text{R}=(x_\text{R},y_\text{R},z_\text{R})^\mathsf{T}\in \mathbb{R}^{3}$, and $\bp_\text{U}=(x_\text{U},y_\text{U},z_\text{U})^\mathsf{T}\in \mathbb{R}^{3}$, respectively. 

\subsection{Channel Model}
We consider the far-field propagation for all the channels, i.e., BS-RIS, RIS-UAV, and BS-UAV channels, which are modeled by taking into consideration the channel parameters, calculated by following the geometrical relationship between any pair of network nodes. For the purpose of simplicity, we only consider the LoS path component for each channel, which is well justified for UAV communications~\cite{Yang2019}. Even though the multi-path components may exist, their sum power is not comparable to the power in the LoS path, especially in millimeter wave (mmWave) frequency bands~\cite{Akdeniz2014}. Characterized by the extended Saleh-Valenzuela channel model~\cite{Ayach2014}, the BS-RIS channel $\bH_\text{B,R} \in \mathbb{C}^{N_\text{R} \times N_\text{B}}$ is modeled as
\begin{equation}\label{channel_H_br}
\bH_\text{B,R} \hspace{-0.1 cm} = \hspace{-0.1 cm}\sqrt{\frac{N_\text{B} N_\text{R}}{\rho_\text{B,R}}} \exp(-j2\pi f_c \tau_\text{B,R})\boldsymbol{\alpha}_\text{R} (\theta^r_\text{B,R}, \phi^r_\text{B,R}) 
\boldsymbol{\alpha}_\text{B}^\mathsf{H} (\phi^t_\text{B,R}- \beta), 
\end{equation}
where $\rho_\text{B,R} \in \mathbb{R}$ is the path loss, dependent on the BS-RIS distance, carrier frequency $f_c$, and shadowing effect, $\tau_\text{B,R}$ is the propagation delay, $\theta^r_\text{B,R} \in \mathbb{R}$, $\phi^r_\text{B,R} \in \mathbb{R}$, and $\phi^t_\text{B,R} \in \mathbb{R}$ are the azimuth, the elevation angle of arrival (AoA), and the elevation angle of departure (AoD), respectively. The channel parameters are calculated based on the centroids of the BS antenna array and the RIS plane, as~\cite{Jiguang2020}  
\begin{align}
x_\text{R} & = x_\text{B} + d_\text{B,R} \cos(\theta^r_\text{B,R}) \cos(\phi^r_\text{B,R}), \\
y_\text{R} & = y_\text{B} + d_\text{B,R} \sin(\theta^r_\text{B,R}) \cos(\phi^r_\text{B,R}), \\
z_\text{R} & = z_\text{B} + d_\text{B,R}  \sin(\phi^r_\text{B,R}),  \\
\phi^t_\text{B,R} &= \phi^r_\text{B,R}, 
\end{align}
where $d_\text{B,R}=\|\bp_\text{B} -\bp_\text{R}\|_2 \in \mathbb{R}$ is the distance between the BS and the RIS.  The normalized array response vectors $ \boldsymbol{\alpha}_\text{R} (\theta^r_\text{B,R}, \phi^r_\text{B,R}) \in \mathbb{C}^{N_\text{R}}$, at the RIS, and $\boldsymbol{\alpha}_\text{B} (\phi^t_\text{B,R}- \beta)  \in \mathbb{C}^{N_\text{B}}$, at the BS, i.e., $\|\boldsymbol{\alpha}_\text{R} (\theta^r_\text{B,R}, \phi^r_\text{B,R})\|_2 = \|\boldsymbol{\alpha}_\text{B}(\phi^t_\text{B,R}- \beta)\|_2 = 1$, are 
\begin{align}\label{alpha}
   & \boldsymbol{\alpha}_\text{R} (\theta^r_\text{B,R}, \phi^r_\text{B,R}) \triangleq \frac{1}{\sqrt{N_\text{R}}}\Big[1, \exp\big(j \frac{2\pi  d_{x}}{\lambda}  \cos(\theta^r_\text{B,R}) \sin(\phi^r_\text{B,R})\big), \nonumber\\
    &\;\;\; \cdots, \exp\big(j \frac{2\pi d_{x}}{\lambda} (N_{\text{R},x} -1) \cos(\theta^r_\text{B,R})\sin(\phi^r_\text{B,R})\big) \Big]^{\mathsf{T}}  \nonumber \\& \otimes\Big[1, \exp\big(j \frac{2\pi  d_{y}}{\lambda}  \sin(\theta^r_\text{B,R}) \sin(\phi^r_\text{B,R})\big), \nonumber\\
    &\;\;\; \cdots, \exp\big(j \frac{2\pi d_{y}}{\lambda} (N_{\text{R},y} -1) \sin(\theta^r_\text{B,R})\sin(\phi^r_\text{B,R})\big) \Big]^{\mathsf{T}},\\
       &\boldsymbol{\alpha}_\text{B}(\phi^t_\text{B,R}- \beta) \triangleq \frac{1}{\sqrt{N_\text{B}}}\Big[1, \exp\big(j \frac{2\pi d_{z}}{\lambda}  \cos(\phi^t_\text{B,R}- \beta) \big), \nonumber\\
    &\;\;\; \cdots, \exp\big(j \frac{2\pi d_{z}}{\lambda} (N_\text{B} -1) \cos(\phi^t_\text{B,R}- \beta)\big) \Big]^{\mathsf{T}},     
 \end{align}
 where $d_x\in \mathbb{R}$, $d_y\in \mathbb{R}$, and $d_z\in \mathbb{R}$ denote the inter-element spacing across $x$, $y$, and $z$ axes, respectively, and $\lambda$ is the wavelength. The effect of orientation angle $\beta \in \mathbb{R}$ is integrated into the array response vector $\boldsymbol{\alpha}_\text{B}( \cdot)$. The RIS-UAV channel $\bH_\text{R,U}\in \mathbb{C}^{N_\text{U} \times N_\text{R}}$ and BS-UAV channel $\bH_\text{B,U}\in \mathbb{C}^{N_\text{U} \times N_\text{B}}$ can be modeled in the same manner, as

 \begin{align}
  \bH_\text{B,U} &= \sqrt{\frac{N_\text{B} N_\text{U}}{\rho_\text{B,U}}} \exp(-j2\pi f_c \tau_\text{B,U})\boldsymbol{\alpha}_\text{U} (\theta^r_\text{B,U}, \phi^r_\text{B,U}) \nonumber\\
  &\times 
\boldsymbol{\alpha}_\text{B}^\mathsf{H} (\phi^t_\text{B,U}- \beta), \\
  \bH_\text{R,U} &= \sqrt{\frac{N_\text{R} N_\text{U}}{\rho_\text{R,U}}} \exp(-j2\pi f_c \tau_\text{R,U}) \boldsymbol{\alpha}_\text{U} (\theta^r_\text{R,U}, \phi^r_\text{R,U})  \nonumber\\
  &\times 
\boldsymbol{\alpha}_\text{R}^\mathsf{H} (\theta^t_\text{R,U},\phi^t_\text{R,U}),  
 \end{align}
where $\boldsymbol{\alpha}_\text{U}( \cdot) \in \mathbb{C}^{N_\text{U}}$ is the array response vector at the UAV, $\rho_\text{B,U}$, $\rho_\text{R,U}$, $\tau_\text{B,U}$, $\tau_\text{R,U}$, $\theta^r_\text{B,U}$, $\phi^r_\text{B,U}$, $\phi^t_\text{B,U}$, $\theta^r_\text{R,U}$, $\phi^r_\text{R,U}$, $\theta^t_\text{R,U}$, and $\phi^t_\text{R,U}$ are defined in the same way as those in~\eqref{channel_H_br}. Since we assume that the antenna plane of the UAV is parallel to that of the RIS, we have $\theta^r_\text{R,U}  = \theta^t_\text{R,U}$ and $\phi^r_\text{R,U}  = \phi^t_\text{R,U}$.

\subsection{LoS Probability}
The LoS probability of the channels relies on the complex environmental propagation conditions, e.g., the height of the buildings, density of the buildings, and spatial distribution of the scatterers. Therefore, different LoS probabilities should be applied for different scenarios, such as urban, suburban, and rural. We, in this work, resort to the International Telecommunication Union Radiocommunication (ITU-R) for the modeling of LoS availability, introduced in~\cite{AlHourani2014,Gapeyenko2021} as 
\begin{equation}\label{p_los}
    p_\text{LoS}^\text{ITU}(h_\text{T},h_\text{R}, \gamma ) \hspace{-0.8mm}=\hspace{-1.3mm} \prod_{m = 0}^M \hspace{-1.3mm}\bigg(1-\exp\bigg(-\frac{\big(h_\text{T} - \frac{(m+0.5)(h_\text{T} - h_\text{R})}{M+1}\big)^2}{2\gamma^2}\bigg)\hspace{-1mm}\bigg),
\end{equation}
where $M = \lfloor r \sqrt{\alpha \kappa}\rfloor -1$ denotes the number of buildings between the pair of network nodes. The variables $r$, $\alpha$, and $\kappa$ denote the ground distance in kilometers between the pair of nodes, the fraction of the area covered by buildings to the total area, and the average number of buildings per unit area, respectively, $\gamma$ is a height distribution parameter, and $h_\text{T}$ and $h_\text{R}$ are the heights for the transmitter and receiver, respectively. By tuning the parameters $\alpha$, $\kappa$, and $\gamma$, we can model the LoS probabilities for all the aforementioned scenarios. 
Thus, we apply the expression~\eqref{p_los} to both of the channels $\bH_\text{B,U}$ and $\bH_\text{R,U}$. In addition, we assume that the RIS is placed in the proximity of the BS, so the LoS availability is always guaranteed. Note that for characterizing $\bH_\text{R,U}$, the RIS is deemed as a virtual transmitter. 

\section{Achievable Rate and Its Average}
\subsection{Achievable Rate}
The benefits of introducing RIS for 3D connectivity are multi-fold: i) When the direct BS-UAV channel exists, tremendous multiplexing gains can be obtained aiming for extremely high data-rate transmission. ii) When the BS-UAV channel is temporally unavailable, the BS can maintain the connectivity via the RIS. We treat the two cases separately in this subsection.
The entire end-to-end channel between the BS and the UAV is summarized as 
\begin{equation}\label{eqn_e2e_H}
\bH = \mathbb{I}(\bH_\text{B,U})\bH_\text{B,U}  +  \mathbb{I}(\bH_\text{R,U})\bH_\text{R,U}\boldsymbol{\Omega}\bH_\text{B,R},  
\end{equation}
where $\mathbb{I}(\cdot) \in \{0,1\}$ is the indicator function; if $\bH_\text{B,U}\neq \mathbf{0}$, $\mathbb{I}(\bH_\text{B,U}) = 1$; otherwise, $\mathbb{I}(\bH_\text{B,U}) = 0$. The same principle is applied to $\mathbb{I}(\bH_\text{R,U})$. The diagonal matrix $\boldsymbol{\Omega} \in \mathbb{C}^{N_\text{R} \times N_\text{R}}$ is the RIS phase control matrix. Unlike the conventional amplify-and-forward (AF) relays, the RIS is supposed to change only the phase shifts of the impinging signals, realizing analog/passive beamforming~\cite{Huangchongwen2019}. Namely, strict constraints are imposed for the diagonal entries of $\boldsymbol{\Omega}$, i.e., $|[\boldsymbol{\Omega}]_{kk}| = 1$, $\forall k \in\{1,2,\cdots, N_\text{R}\}$. By referring to~\eqref{p_los}, the LoS probabilities of $\bH_\text{B,U}$ and $\bH_\text{R,U}$ are 
\begin{align}
    \text{Pr}(\mathbb{I}(&\bH_\text{B,U}) \hspace{-0.7mm}=\hspace{-0.7mm} 1)\hspace{-0.7mm} =\nonumber\\
 &   \hspace{-1mm} \prod_{m = 0}^M\hspace{-1.5mm} \bigg(1-\exp\bigg(-\frac{\big(z_\text{B} - \frac{(m+0.5)(z_\text{B} - z_\text{U})}{M+1}\big)^2}{2\gamma^2}\bigg)\hspace{-1mm}\bigg), \\
\text{Pr}(&\mathbb{I}(\bH_\text{R,U}) \hspace{-0.7mm}=\hspace{-0.7mm} 1)\hspace{-0.7mm} =\nonumber\\
&\hspace{-1mm}\prod_{m = 0}^M\hspace{-1.5mm} \bigg(1-\exp\bigg(-\frac{\big(z_\text{R} - \frac{(m+0.5)(z_\text{R} - z_\text{U})}{M+1}\big)^2}{2\gamma^2}\bigg)\hspace{-1mm}\bigg). 
\end{align}

\begin{figure*}[t!]
\begin{equation} \label{achi_rate}
R =  
   \begin{cases}
        \log_2 \Big(\det\big( \bI + \frac{P}{\sigma^2 }\bW^\mathsf{H}\bH \bF \bF^\mathsf{H} \bH^\mathsf{H} \bW\big)\Big), & \text{if } \mathbb{I}(\bH_\text{B,U}) =1, \mathbb{I}(\bH_\text{R,U}) =  1,\\
        \log_2 \Big(\det\big( 1 + \frac{P}{\sigma^2 }\bw^\mathsf{H}\bH_\text{B,U} \bf \bf^\mathsf{H} \bH_\text{B,U}^\mathsf{H} \bw\big)\Big), & \text{if } \mathbb{I}(\bH_\text{B,U}) =1, \mathbb{I}(\bH_\text{R,U}) =  0,\\
         \log_2 \Big(\det\big( 1 + \frac{P}{\sigma^2 }\bw^\mathsf{H}\bH_\text{R,U}\boldsymbol{\Omega}\bH_\text{B,R} \bf \bf^\mathsf{H} \bH_\text{B,R}^\mathsf{H} \bH_\text{R,U}^\mathsf{H}\boldsymbol{\Omega}^\mathsf{H} \bw\big)\Big), & \text{if } \mathbb{I}(\bH_\text{B,U}) =0, \mathbb{I}(\bH_\text{R,U}) =  1,\\
         0, & \text{if } \mathbb{I}(\bH_\text{B,U}) =0, \mathbb{I}(\bH_\text{R,U}) =  0.
    \end{cases}
\end{equation}

\hrulefill

\end{figure*}

The achievable rate between the BS and the UAV depends on the LoS availability of $\bH_\text{B,U}$ and $\bH_\text{R,U}$. Thus, a segmented function is considered for the achievable rate in bps/Hz, as shown in~\eqref{achi_rate} on the top of the next page. For case i) $\mathbb{I}(\bH_\text{B,U}) =\mathbb{I}(\bH_\text{R,U}) =  1$, it also depends on the design of the combining matrix $\bW \in \mathbb{C}^{N_\text{U} \times 2}$, the precoder $\bF \in \mathbb{C}^{N_\text{B} \times 2}$, and $\boldsymbol{\Omega}$. For case ii) $\mathbb{I}(\bH_\text{B,U}) =1, \mathbb{I}(\bH_\text{R,U}) =  0$, it also depends on the design of the combining vector $\bw\in \mathbb{C}^{N_\text{U}}$ and the beamforming vector $\bf\in \mathbb{C}^{N_\text{B}}$. For the case iii) $\mathbb{I}(\bH_\text{B,U}) =0, \mathbb{I}(\bH_\text{R,U}) =  1$, it also depends on the design of $\bw$, $\bf$, and $\boldsymbol{\Omega}$. The notations $P$ and $\sigma^2$ denote the  transmit power at the BS and the noise variance at the UAV.  

\subsubsection{Case i) $\mathbb{I}(\bH_\textsc{B,U}) =\mathbb{I}(\bH_\textsc{R,U}) =  1$}
In this case, spatial multiplexing gains can be achieved, since two different routes from the BS to the UAV can be concurrently available with the end-to-end channel $\bH = \bH_\text{B,U}+\bH_\text{R,U}\boldsymbol{\Omega}\bH_\text{B,R}$, where $\bH_\text{B,U} \neq \mathbf{0}$ and $\bH_\text{R,U} \neq \mathbf{0}$. We focus on the design of $\bW$, $\bF$, and $\boldsymbol{\Omega}$ as to maximize the achievable rate. The optimization problem is formulated as~\cite{Renwang2023}
\begin{subequations}
\begin{align} \label{eq_rate_opt}
\mathcal{P}_1:   \max\limits_{\bW,\bF, \boldsymbol{\Omega }} \;\;  \;&\log_2 \Big(\det\big( \bI_2 + \frac{P}{\sigma^2 }\bW^\mathsf{H}\bH \bF \bF^\mathsf{H} \bH^\mathsf{H} \bW\big)\Big), \\
    \textit{s.t.}\;\; 
    &\mathrm{Tr}(\bW^\mathsf{H} \bW) = 1, \label{W_constraint}\\
    &\mathrm{Tr}(\bF^\mathsf{H} \bF) = 1, \label{F_constraint}\\
    &|[\boldsymbol{\Omega}]_{kk}| = 1, \forall k, \label{RIS_constraint}
    \end{align}
\end{subequations}
which is non-convex due to the non-convex constraint in~\eqref{RIS_constraint}. Therefore, it is challenging to find the optimal solution. In order to ease the analysis that follows, we use a simplified yet intuitive two-stage approach to sequentially optimize $ \boldsymbol{\Omega}$ and $\{\bW, \bF\}$. We first focus on the term $\bH_\text{R,U}\boldsymbol{\Omega}\bH_\text{B,R}$, and define $\eta = \sqrt{\frac{N_\text{R} N_\text{U}}{\rho_\text{R,U}}}\sqrt{\frac{N_\text{B} N_\text{R}}{\rho_\text{B,R}}} \exp(-j2\pi f_c \tau_\text{R,U})\exp(-j2\pi f_c \tau_\text{B,R})
\boldsymbol{\alpha}_\text{R}^\mathsf{H} (\theta^t_\text{R,U},\phi^t_\text{R,U}) \\\boldsymbol{\Omega} \boldsymbol{\alpha}_\text{R} (\theta^r_\text{B,R}, \phi^r_\text{B,R})$. In this regard, $\bH$ in~\eqref{eqn_e2e_H} can be reformulated as \begin{equation}
    \bH(\eta) = \underbrace{\bH_\text{B,U}}_\text{Rank One} + \underbrace{\eta \boldsymbol{\alpha}_\text{U} (\theta^r_\text{R,U}, \phi^r_\text{R,U})\boldsymbol{\alpha}_\text{B}^\mathsf{H} (\phi^t_\text{B,R}- \beta) }_\text{Rank One},
\end{equation}
which is a summation of two rank-one matrices. Here, we replace $\bH$ as $\bH(\eta)$ to show its dependence on $\eta$. Based on the following facts:
\begin{equation}\label{eqn_lim_alpha_B}
    \lim_{N_\text{B} \rightarrow \infty} \boldsymbol{\alpha}_\text{B}^\mathsf{H} (\phi^t_\text{B,U}- \beta) \boldsymbol{\alpha}_\text{B} (\phi^t_\text{B,R}- \beta) \approx 0, \; \text{if } \;\phi^t_\text{B,U} \neq \phi^t_\text{B,R}, 
    \end{equation}
    \begin{align}
     \lim_{N_\text{U} \rightarrow \infty} \boldsymbol{\alpha}^\mathsf{H}_\text{U} (\theta^r_\text{B,U}, \phi^r_\text{B,U}) \boldsymbol{\alpha}_\text{U} (\theta^r_\text{R,U}, \phi^r_\text{R,U})  \approx 0, \; \text{if}\; &\theta^r_\text{R,U} \neq \theta^r_\text{B,U}\nonumber\\
      & \&\phi^r_\text{R,U} \neq \phi^r_\text{B,U}, \label{eqn_lim_alpha_U}
\end{align}
which can be readily proven. Thus, the singular value decomposition (SVD) of $\bH(\eta)$ is approximated by 
\begin{equation}
    \bH(\eta) \approx \bU \mathrm{diag}([\sqrt{N_\text{B} N_\text{U}/\rho_\text{B,U}}  \;\; |\eta|]) \bV^\mathsf{H},
\end{equation}
where $\bU = [ \exp(-j2\pi f_c \tau_\text{B,U})\boldsymbol{\alpha}_\text{U} (\theta^r_\text{B,U}, \phi^r_\text{B,U}) ;  \boldsymbol{\alpha}_\text{U} (\theta^r_\text{R,U}, \phi^r_\text{R,U}) ] $ and $\bV = [\boldsymbol{\alpha}_\text{B} (\phi^t_\text{B,U}- \beta) ;  \boldsymbol{\alpha}_\text{B} (\phi^t_\text{B,R}- \beta) ]$. According to~\eqref{eqn_lim_alpha_B} and~\eqref{eqn_lim_alpha_U}, $\bU$ and $\bV$ are nearly partial of unitary matrices, i.e., $\bU^\mathsf{H}\bU \approx  \bI_2$  and $\bV^\mathsf{H}\bV \approx  \bI_2$. 

To this end, we first maximize $|\eta|$ in the first stage, which is equal to $\sqrt{\frac{N_\text{R} N_\text{U}}{\rho_\text{R,U}}}\sqrt{\frac{N_\text{B} N_\text{R}}{\rho_\text{B,R}}}$. Afterwards, we perform SVD of $\bH(\eta = \sqrt{\frac{N_\text{R} N_\text{U}}{\rho_\text{R,U}}}\sqrt{\frac{N_\text{B} N_\text{R}}{\rho_\text{B,R}}})$ to design the optimal $\bW$ and $\bF$ in the second stage. The optimal closed-form solutions for $\boldsymbol{\Omega}$, $\bW$, and $\bF$ are summarized as 
\begin{align}
    [\boldsymbol{\Omega}]_{kk} &=\exp\bigg\{j\Big( \phase{[\boldsymbol{\alpha}_\text{R} (\theta^t_\text{R,U},\phi^t_\text{R,U})]_k } - \phase{  [\boldsymbol{\alpha}_\text{R} (\theta^r_\text{B,R}, \phi^r_\text{B,R})]_k}\Big)\bigg\} \nonumber\\
    &\times \exp\big(j2\pi f_c (\tau_\text{R,U}+\tau_\text{B,R})\big),\\
    \bW &= \frac{\sqrt{2}}{2}\bU, \bF = \frac{\sqrt{2}}{2}\bV,
\end{align}
where $\bW$ and $\bF$ meet the constraints specified in~\eqref{W_constraint} and~\eqref{F_constraint}. 

The achievable rate, rewritten as $R_1$, for this case is 
\begin{equation}
    R_1 \approx   \log_2 \Big(1+ \frac{P}{4 \sigma^2} \frac{N_\text{B} N_\text{U}}{\rho_\text{B,U}} \Big) + \log_2 \Big(1+ \frac{P}{4 \sigma^2} \frac{N_\text{R} N_\text{U}}{\rho_\text{R,U}} \frac{N_\text{B} N_\text{R}}{\rho_\text{B,R}}  \Big).
\end{equation}

\subsubsection{Case ii) $\mathbb{I}(\bH_\textsc{B,U}) =1, \mathbb{I}(\bH_\textsc{R,U}) =  0$} For this case, we only need to perform SVD of $\bH_\text{B,U}$ to design the optimal beamforming vectors $\bw$ and $\bf$. The calculation of the achievable rate is straightforward. The achievable rate, redefined as $R_2$, for this case is $R_2 = \log_2 \Big(1+ \frac{P}{\sigma^2} \frac{N_\text{B} N_\text{U}}{\rho_\text{B,U}} \Big)$.

\subsubsection{Case iii) $\mathbb{I}(\bH_\textsc{B,U}) =0, \mathbb{I}(\bH_\textsc{R,U}) = 1$} For this case, we can follow the two-stage approach, already applied for case i). The achievable rate, redefined as $R_3$, for this case is $R_3 = \log_2 \Big(1+ \frac{P}{ \sigma^2} \frac{N_\text{R} N_\text{U}}{\rho_\text{R,U}} \frac{N_\text{B} N_\text{R}}{\rho_\text{B,R}}  \Big)$.

\subsection{Average Achievable Rate}
We take the LoS probabilities and their associated achievable rates into consideration and calculate the average achievable rate as 
\begin{align}
    \bar{R} &= R_1 \text{Pr}(\mathbb{I}(\bH_\text{B,U}) = 1) \text{Pr}(\mathbb{I}(\bH_\text{R,U})=1) \nonumber\\
    &+ R_2 \text{Pr}(\mathbb{I}(\bH_\text{B,U}) = 1) \text{Pr}(\mathbb{I}(\bH_\text{R,U})=0) \nonumber\\&+R_3 \text{Pr}(\mathbb{I}(\bH_\text{B,U}) = 0) \text{Pr}(\mathbb{I}(\bH_\text{R,U})=1). 
\end{align}


\section{Simulation Results}\label{sec:results}
\begin{figure}[t]
	\centering
\includegraphics[width=.9\linewidth]{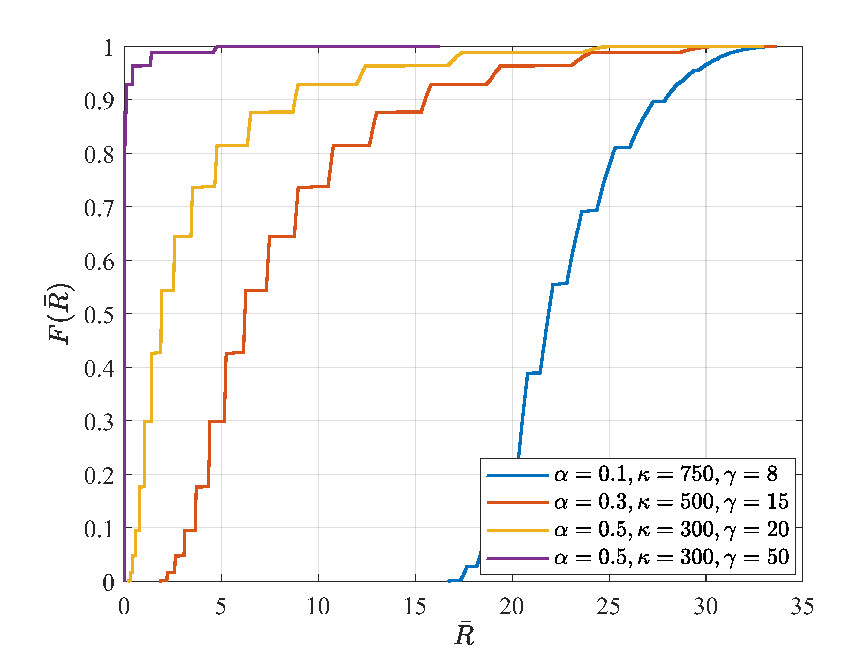}
	\caption{CDFs of average achievable rate for different environments, i.e., suburban, urban, dense urban, and highrise urban, with $f_c = 28$ GHz.}
		\label{CDF_Avg_Rate}
\end{figure}

In this section, we study the average achievable rate for an RIS-assisted B5G network and compare it with its RIS-free counterpart. For simplicity, we only consider the free-space path loss, which is modeled as: $\rho = d^2 f_c^2 / 10^{8.755}$, where $f_c$ (in KHz) is the carrier frequency, defined as $f_c = \frac{c}{\lambda}$ with $c$ being the speed of light ($3\times 10^8$ m/s). We omit the subscripts of $\rho$ and $d$ to make them general and applicable to all the path losses and distances. The system parameters are set as: $f_c \in \{28, 3.5\}$ GHz, $\beta = \pi/3$, $N_\text{B} = 8 \times 8$, $N_\text{R} = 20 \times 20$, $N_\text{U} = 8 \times 8$, $\bp_\text{B} = (0, 0, 10)$, $\bp_\text{R} = (0.5, 0.5, 9.5)$, and bandwidth $B = 20$ MHz. The UAV is supposed to be located at any possible position with $x_\text{U}>0, y_\text{U}>0, z_\text{U} > 0$. To model the LoS probability, we follow~\cite{AlHourani2014} for the setup for $(\alpha, \kappa, \gamma)$:  suburban (0.1, 750, 8),
urban (0.3, 500, 15), dense urban (0.5, 300, 20), and highrise
urban (0.5, 300, 50). We fix the height for the UAV as $100$ meters, i.e., $z_\text{U} = 100$, while its $x$-axis and $y$-axis are uniformly distributed within $[100, 1000]$ meters when $f_c = 28$ GHz and $[100, 2000]$ meters when $f_c = 3.5$ GHz. The cumulative distribution functions (CDFs) of the average achievable rate are shown in Fig.~\ref{CDF_Avg_Rate} for different propagation environments with $f_c = 28$ GHz. It can be seen from the simulation curves that the higher LoS availability (e.g., suburban) the better the average achievable rate. By following $B = 20$ MHz, the BS can offer up to several hundred Mbps data rate for the flying UAVs in the suburban scenario.   

\begin{figure}[t]
	\centering
\includegraphics[width=1.12\linewidth]{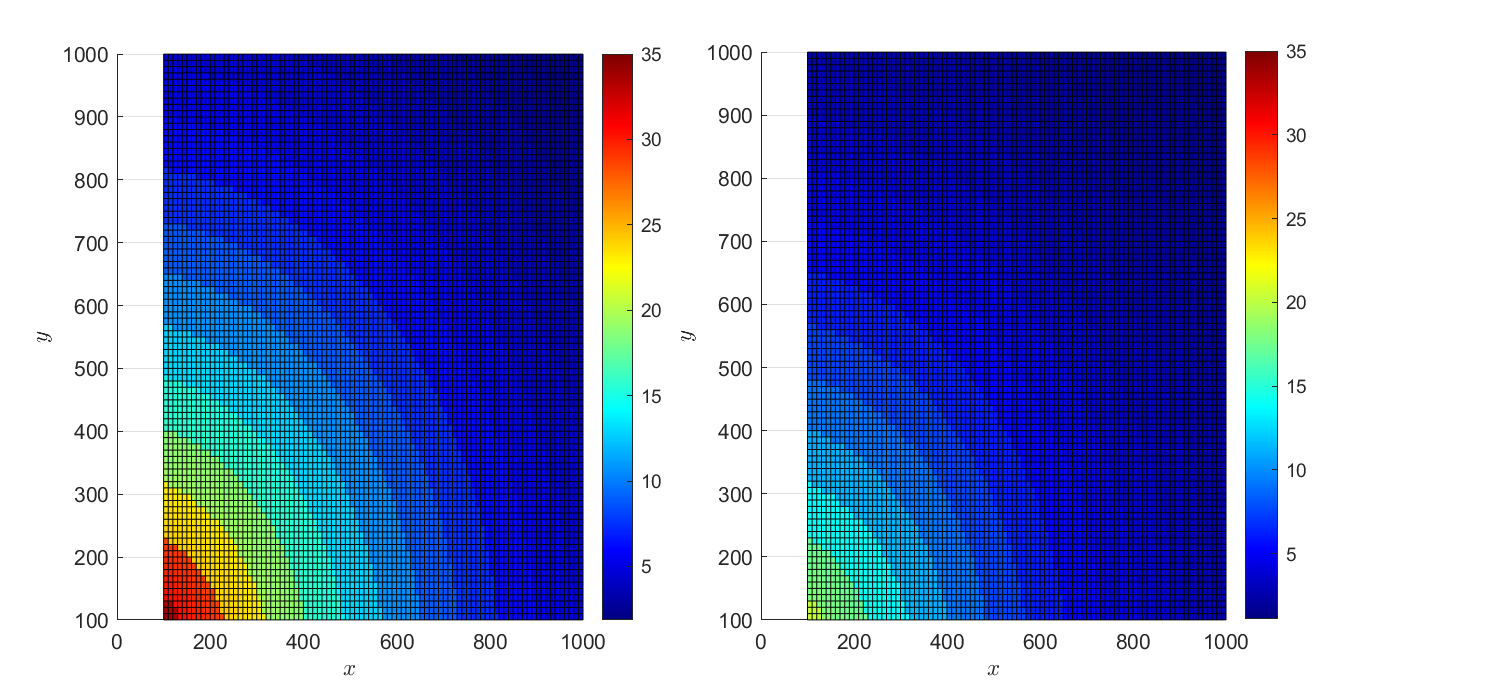}
	\caption{Comparison between RIS-assisted (left) and RIS-free (right) UAV communication networks in terms of average achievable rate at 28 GHz (mmWave). The urban scenario is evaluated.}
		\label{Heat_map_28}
\end{figure}

\begin{figure}[t]
	\centering
\includegraphics[width=1\linewidth]{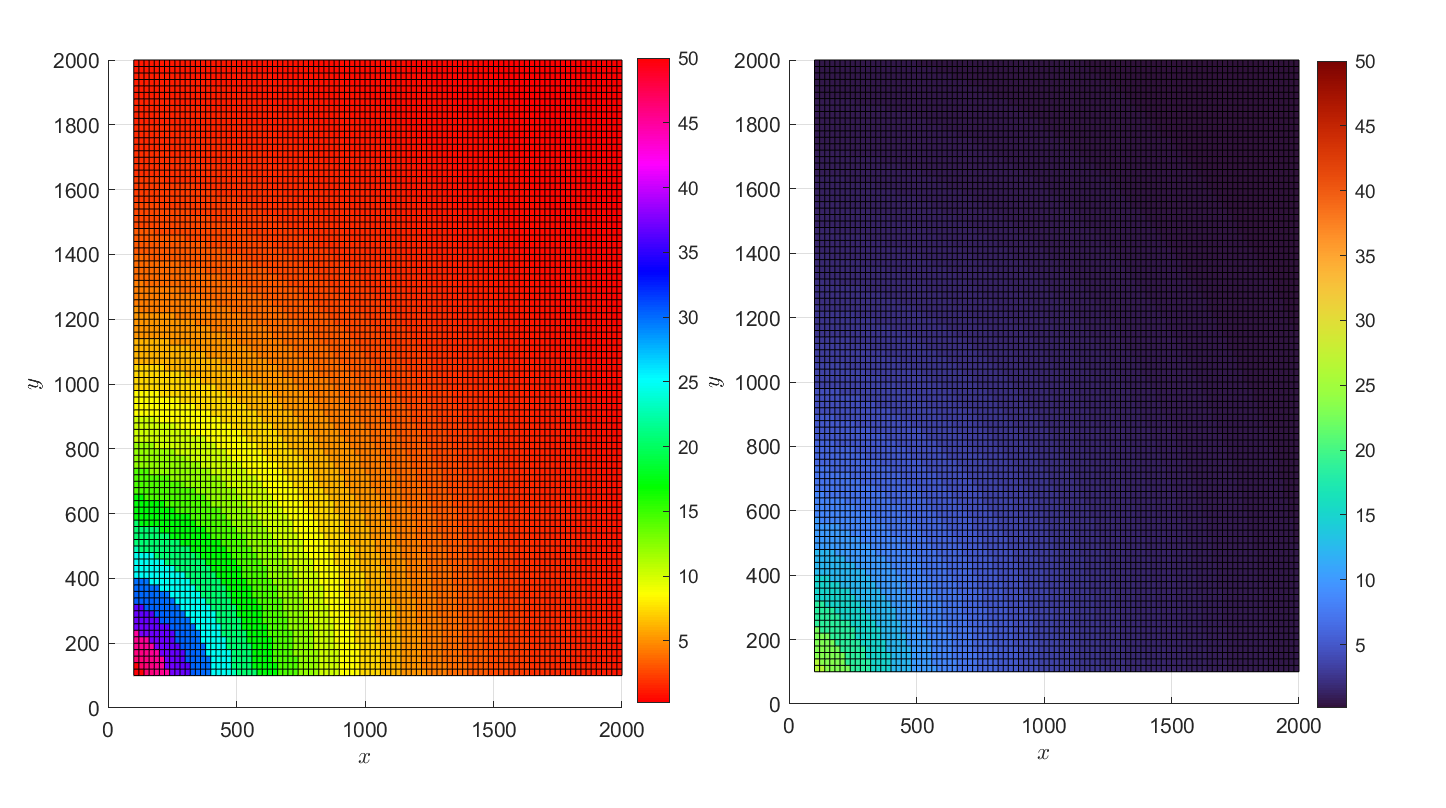}
	\caption{Comparison between RIS-assisted (left) and RIS-free (right) UAV communication networks in terms of average achievable rate at 3.5 GHz (C band). The urban scenario is evaluated.}
		\label{Heat_map_3dot5}
\end{figure}

Comparisons of the average achievable rates of the RIS-assisted and RIS-free UAV communication networks are shown in Figs.~\ref{Heat_map_28} and~\ref{Heat_map_3dot5} for both carrier frequencies, i.e., mmWave and C band. As we can see, significant rate gains can be obtained by  introducing the RIS to the existing 5G cellular networks, confirming the 3D connectivity enhancements that can be unleashed by RISs to UAVs. It also becomes clear that higher frequencies offer a shorter communication range due to more server path loss.

\section{Conclusions}\label{sec:conclusion}
In this paper, we have studied the 3D connectivity enabled by exploiting the existing 5G BS together with an RIS. We have investigated the average achievable rate while taking into consideration of the LoS probabilities under different propagation environments. Simulation results have shown that with the introduction of RIS, substantially higher average achievable rates can be obtained compared to its RIS-free counterpart. 

\section*{Acknowledgement}
This contribution of Aymen Fakhreddine has been partly funded by FWF -- Der Wissenschaftsfonds (Austrian Science Fund) ESPRIT program under grant number ESP-54.

\bibliographystyle{IEEEtran}
\bibliography{biblio,bettstetter}

\end{document}